\documentclass[aps,pre,reprint,twocolumn,amsmath,amssymb,groupedaddress]{revtex4-2}

\usepackage{graphicx}
\usepackage[hidelinks]{hyperref}
\usepackage[]{units}

\usepackage{comment}

\usepackage{verbatim}
\usepackage{color}

\begin{document}

\title{Coat stiffening can explain invagination of clathrin-coated membranes}

\date{\today}
\author{Felix Frey$^{1}$}
\author{Ulrich S. Schwarz$^{2}$}
\email{To whom correspondence should be addressed; E-mail: schwarz@thphys.uni-heidelberg.de}
\affiliation{$^{1}$Institute of Science and Technology Austria, 3400 Klosterneuburg, Austria}
\affiliation{$^{2}$Institute for Theoretical Physics and BioQuant, Heidelberg University, 69120 Heidelberg, Germany}

\begin{abstract}
Clathrin-mediated endocytosis is the main pathway used by eukaryotic cells to take up extracellular material, but the dominant physical mechanisms driving this process are still elusive.
Recently several high-resolution imaging techniques have been used on different cell lines to measure the geometrical properties of clathrin-coated pits over their whole lifetime. 
Here we first show that the combination of all datasets with 
the recently introduced cooperative curvature model defines a consensus pathway,
which is characterized by a flat-to-curved transition at finite area, followed by linear growth and subsequent saturation of curvature. 
We then apply an energetic model for the composite of plasma membrane and clathrin coat to this consensus pathway to show that the dominant mechanism for invagination could be coat stiffening, which might originate from cooperative interactions between the different clathrin molecules and progressively drives the system towards its intrinsic curvature. 
Our theory predicts that two length scales determine the invagination pathway, namely the patch size at which the flat-to-curved transition occurs and the final pit radius.
\end{abstract}

\maketitle

\section{Introduction}

Each biological cell is defined and protected by a plasma membrane, but
also needs to transport nutrients or signaling molecules across it \cite{alberts2015}.
To take up extracellular material, eukaryotic 
cells have developed different uptake strategies
\cite{zhang2015}. Among these, clathrin-mediated endocytosis
(CME) is the main uptake route and used mainly
for particles in the range from 20 to 300 nm, which
also includes many common viruses \cite{kaksonen2018}.
In order to bend the plasma membrane inwards and to form 
a transport vesicle, a hexagonal lattice made from the protein clathrin
is assembled at the plasma membrane, which is punctuated
by a few pentagons to generate curvature \cite{PICCO2018}.
These networks are naturally formed by clathrin molecules, because they have the form of triskelia after assembly of three heavy chains, each decorated with one light chain. 
Clathrin triskelia feature multiple binding sites to each other as well as
an intrinsic curvature \cite{fotin2004}. 
Remarkably, clathrin triskelia can self-assemble without any
additional factors into closed cages that resemble fullerenes,
with twelve pentagons in a sea of hexagons \cite{KIRCHHAUSEN1981}.
In reconstitution assays with lipid vesicles,
clathrin binds to the membrane through
adapter proteins like AP180 and then deforms it \cite{saleem2015}.
In the cell, the situation under which clathrin lattices assemble is more complex, since here many more proteins participate in the assembly and  invagination process \cite{smith2022}. 
In particular, an actin network can form around the growing pit, that can pull and push the invaginating membrane inwards \cite{serwas2022,yang2022}.

The exact time point at which curvature is generated during CME has been debated since many decades \cite{wood2021,chen2020}. 
Early work with cells supported the constant area model (CAM), in which clathrin triskelia first grow into large flat patches, which then invaginate.
In contrast, reconstitution experiments with lipids and clathrin tend to favor the constant curvature model (CCM), in which curvature is generated right from the start of the clathrin assembly \cite{saleem2015}. 
Today it is accepted that the actual process is a mixture of both scenarios, with both area and curvature growing in time during CME.

Recently different high-resolution imaging techniques have been used to gain insight into the spatiotemporal coordination of CME in mammalian cells \cite{avinoam2015,bucher2018,Scott2018,mund2023}.
An early study with electron tomography (ET) has demonstrated that the curvature of the clathrin coats is not constant, but increases during invagination \cite{avinoam2015}. 
Using high-speed atomic force microscopy (HS-AFM) for imaging and nano-dissection, it has been found that clathrin lattices relax after cutting and thus store elastic energy \cite{tagiltsev2021}. 
With super-resolution microscopy (SRM), it was found that the curvature of the clathrin coat starts to be generated only once around half of the final clathrin coat has been assembled in flat patches and that later it saturates \cite{mund2023}. 
What all these techniques have in common is that snapshots are taken of clathrin coats that can be sorted according to the state of invagination, where usually the invagination angle is taken as a surrogate for time.
However, until now these different datasets have not been compared with each other. 
In particular, the HS-AFM data has not yet been explicitly evaluated as a function of pseudo-time.

\begin{figure*}
\centering
\includegraphics[width=.85\textwidth]{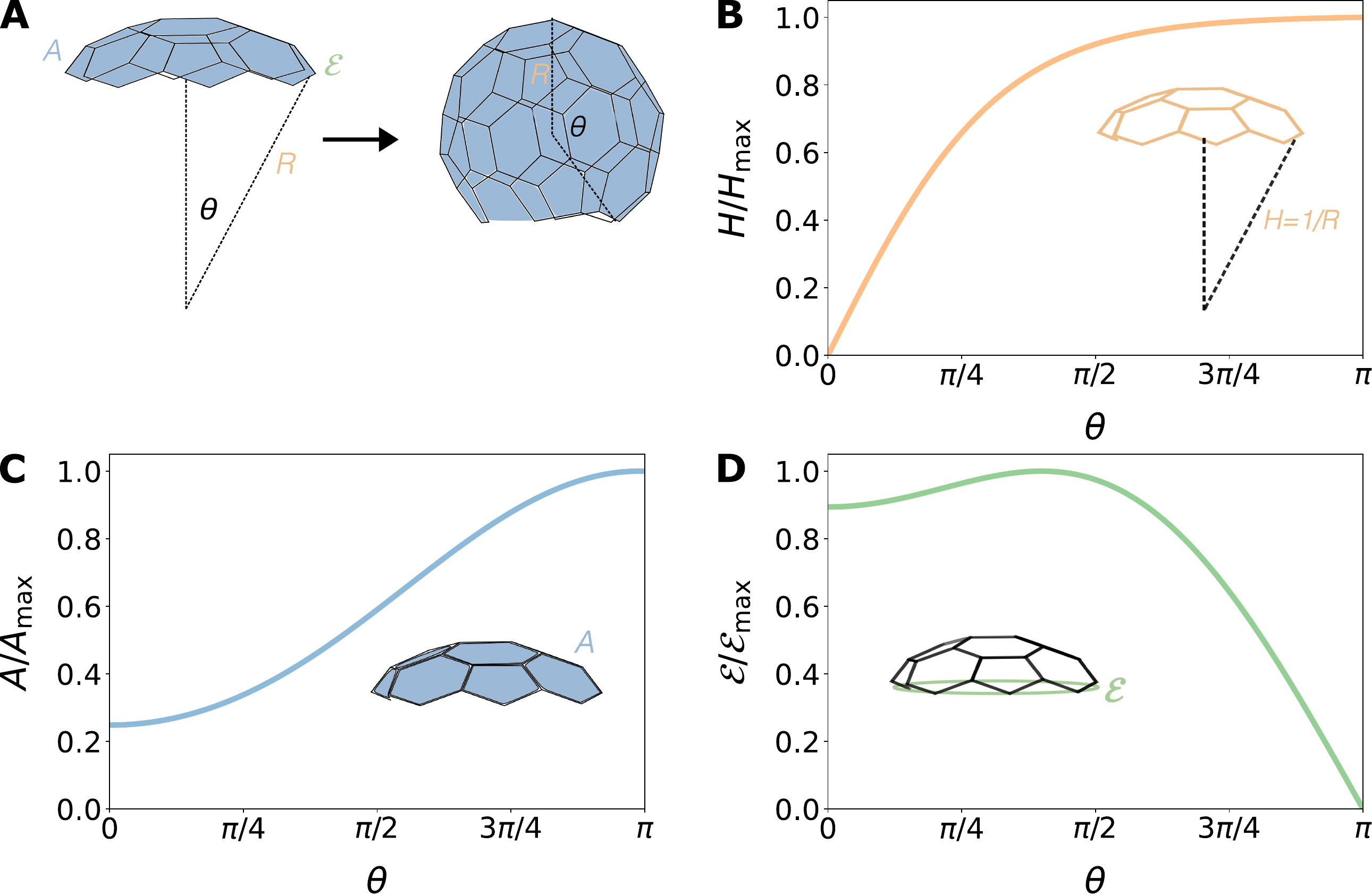}
\caption{\textbf{Cooperative curvature model (CoopCM).} (A) The clathrin coat is assumed to have the shape of a spherical cap with invagination angle $\theta$, cap radius $R$, area $A$ and edge length $\mathcal{E}$.
Curvature $H=1/R$ is therefore the same everywhere, but changes in time, as does the invagination angle $\theta$.
(B) Curvature $H$, (C) area $A$ and (D) edge length$\mathcal{E}$, as predicted by the CoopCM, plotted against the invagination angle $\theta$ and normalized to their maxima (for $H_0=\gamma$). 
The invagination angle $\theta$ acts as surrogate for time.
}
\label{fig:Fig1}
\end{figure*}

Although their molecular architecture suggests that individual clathrin triskelia can independently generate coat curvature, in recent years it has become increasingly clear that coat invagination arises from the collective interaction of clathrin triskelia within the clathrin lattice.
In particular, it has been shown that invagination is favored by exchange of the clathrin light chains \cite{maib_cargo_2018}, that clathrin assembly is facilitated on curved membranes \cite{dannhauser_effect_2015,zeno2021,Cail2022} and that clathrin triskelia undergo conformational switches that increase curvature \cite{obashi_conformational_2023}.
Moreover, dissection of the clathrin coat using AFM has demonstrated that the assembled lattice is elastically frustrated \cite{tagiltsev2021}.
Recent experimental studies also highlighted the role of actin during coat invagination \cite{serwas2022}.
In particular, it has been suggested that in mammalian cells, polymerizing actin could drive the invagination of the clathrin coat by pushing on its edges \cite{yang2022}.
Moreover, it has been discussed that coat rearrangements \cite{sochacki2020} or lattice vacancies \cite{frey2020} could facilitate the invagination process.

To provide a simple mathematical description of coat assembly and invagination, we have recently suggested a conceptually transparent and analytically solvable kinetic model, the cooperative curvature model (CoopCM), that agrees with the growth kinetics observed with SRM \cite{mund2023}.
The CoopCM is  based on only a few simple assumptions that are motivated by the experimental observations, namely that the growing pit resembles a spherical cap, that area growth occurs mainly at the rim and that curvature is generated everywhere in the lattice, but in a cooperative manner, until a preferred value is reached. 
Although this phenomenological approach is effective, it does not address the underlying mechanisms.
In principle, different physical mechanisms can contribute to drive coat invagination, including generation of spontaneous curvature, coat stiffening and line tension \cite{frey2020b,tagiltsev2021}. 

In this work, we first discuss the CoopCM in more detail and show explicitly why it predicts a flat-to-curved transition at finite coat area.
We then combine experimental data from different high-resolution techniques and different mammalian cell lines with the CoopCM to define a consensus pathway of CME. 
We next combine this consensus pathway with an energetic model for the composite of  plasma membrane and clathrin coat.
Within the framework of this model, we find that dynamic coat stiffening is a simple way in which invagination can be achieved. 
Because in the interface model invagination is driven by spontaneous curvature and stiffness penalizes deviations from this spontaneous curvature, high stiffness is required to stabilize the target state, but generates a prohibitively large energy penalty for the initially flat clathrin coat. 
Therefore, a gradual increase of stiffness could balance these two requirements of the system.

\section{The cooperative curvature model}

Previously, we showed that the kinetics of the invagination of the clathrin coat is described by the phenomenological CoopCM, which is justified as it effectively fits the invagination data of clathrin coats \cite{mund2023}.
In the following, we briefly summarize the model and the underlying main assumptions. We also demonstrate
that the CoopCM can be derived with both
phenomenological and geometrical arguments.

Our first main assumption is that the clathrin coat takes the shape of a spherical cap. This assumption
is supported well by electron microscopy
\cite{avinoam2015} as well as by SRM \cite{Wu2023,mund2023} for mammalian cell types
(but not for yeast, which in addition has
a high turgor pressure). A spherical cap
is characterized by only two quantities, namely the invagination angle $\theta$ and the cap radius $R$ (Fig.~\ref{fig:Fig1}A).
The two most important geometrical quantities for kinetic and energetic models are the cap area $A$ and the edge length $\mathcal{E}$, which for the spherical cap read $A=2 \pi R^2 (1-\cos \theta)$ and $\mathcal{E}=2 \pi R \sin \theta$, respectively. 
It is important to note that the formula for the area of the spherical cap includes the flat circular disk in the double limit when $\theta \rightarrow 0$ and at the same time $R \rightarrow \infty$ with the disk radius $R \sin \theta \approx R \theta$ remaining finite.

Instead of giving a fully dynamic description, we describe how the coat curvature $H=1/R$ evolves with $\theta$, which we use as a surrogate of time.
The reason for this is that experimentally only snapshots of clathrin coats are taken, which, assuming that coat invagination is an irreversible process, can be ordered according to their opening angle $\theta$.
Within the spherical cap assumption, 
curvature than is the second important variable.
Experimentally it has been found that
curvature changes the fastest at the
beginning of invagination \cite{mund2023}.
We therefore assume that curvature $H$ is the dominant fast dynamic variable and that coat area $A$ is controlled by an independent growth mechanism along the edge of the clathrin coat. The initial growth 
rate is called $\gamma$ and should be determined
by the interaction of the clathrin triskelia,
which self-assembly into a hexagonal lattice with
pentagonal defects.

As the clathrin coat invaginates, the generation of coat curvature has to slow down eventually, because the clathrin coat approaches the final coat curvature $H_0 = 1 / R_0$, which is set by the curvature of the clathrin triskelia, interactions with other clathrin triskelia and interactions with adaptor and accessory proteins at the membrane \cite{zeno2021,sochacki2020}.
This coat curvature $H_0$ is expected to be smaller than the curvature of clathrin cages $H_\mathrm{c}=1/R_\mathrm{c}$, which is determined solely by the curvature of the clathrin triskelia and the interactions between the other clathrin triskelia \cite{fotin2004,wood2021}.
The cage radius $R_\mathrm{c}$ has been measured to be in the range $\unit[32.5-50]{nm}$ \cite{saleem2015,hassinger2017}.
For typical pit sizes (typical membrane radius \unit[40]{nm} \cite{avinoam2015} plus \unit[15]{nm} accounting for the thickness of the clathrin coat and its gap to the membrane \cite{jin2006,saleem2015})
the expected value for $R_0$ is around $\unit[55]{nm}$, which indeed is above the range for radii for the cages.

As the coat reaches the saturation curvature $H_0$, the coat curvature will stop increasing and a stable steady state of invagination emerges. 
Therefore, this curvature acts as a control mechanism for the invagination and effectively sets the pit size.
Since the curvature saturation only sets in at the late stage of the invagination process, we assume that it is proportional to $H^2$. 
Importantly, the choice of $H^2$, rather than of $H$, is also justified phenomenologically, as it gives better fits \cite{mund2023}. 
Although $H$ is a high-level feature of the system, it is tempting to speculate that the functional form of $H^2$ also indicates that curvature saturation is a signature of cooperativity between clathrin triskelia.

Together, we now have a simple evolution equation for the coat curvature as a function of the invagination angle 
\begin{equation}
\frac{\mathrm{d} H}{\mathrm{d} \theta}= \gamma \left ( 1 - \frac{H^2}{H_0^2} \right ) \, .
\label{eq:ODE_H_2}
\end{equation} 
With the initial condition $H(\theta=0)=0$ 
the solution to Eq.~(\ref{eq:ODE_H_2}) is
\begin{equation}
H(\theta)=H_0 \tanh \left( \frac{\gamma}{H_0} 
\theta \right) \, .
\label{eq:solution_H_2}
\end{equation}
Consequently, the radius of the coat is defined as the inverse of the coat curvature$R(\theta)=1/ H(\theta)$, where $R_0=1/H_0$.

Besides this phenomenological derivation, we want to demonstrate that Eq.~(\ref{eq:ODE_H_2}) can be also motivated using geometrical arguments.
Earlier we have shown that the CoopCM can be formulated also as a fully dynamic description that predicts the characteristic square root dependence of curvature with time that was observed experimentally \cite{mund2023}. 
Until Section~\ref{sec:dynamic_coat}, we do not address time explicitly, because we first focus on the geometry of the system, which is sufficient to develop an energetic description. 
However, as done earlier for the fully dynamical version of the CoopCM, we now make the assumption that area growth along the edge occurs independently of coat invagination, which includes the possibility that invagination starts at finite area.

Assuming that the clathrin coat takes the shape of a spherical cap, defined by the coat radius $R$ (or its inverse the coat curvature $H$) and opening angle $\theta$, we can solve the area formula for the coat curvature. 
As a result we obtain $H=\sqrt{2 \pi (1- \cos \theta)/A}$, which is justified as long as $A > 0$. 
Expanding the coat curvature $H$ for a patch with finite area ($A > 0$) in $\theta$ around the flat state ($\theta = 0$) yields 
\begin{equation}
H= \sqrt{\frac{\pi}{A} } \left(\theta-\frac{\theta^3}{24} \right) + \mathcal{O}(\theta^5) \, .
\label{eq:exapnsion_H}
\end{equation}
The expansion is justified as long as $A$ changes on another (slower) time scale. 

In order to obtain a differential equation for the coat curvature, we assume that the derivative of coat curvature $\mathrm{d} H/\mathrm{d} \theta$ takes the form of a polynomial in $H$. 
To include non-linear effects like cooperativity between clathrin triskelia we consider the equation to be of second order in $H$, $\mathrm{d} H /{\mathrm{d} \theta}=c_0-c_1 H -c_2 H^2$, where $c_0, c_1$ and $c_2$ are expansion coefficients. 
We next use the expression for $H$ of Eq.~(\ref{eq:exapnsion_H}) on the polynomial and compare the resulting equation with the derivative of Eq.~(\ref{eq:exapnsion_H}) with respect to $\theta$.
From the comparison of the two equations, we find that $c_1 = 0$ since the derivative of Eq.~(\ref{eq:exapnsion_H}) does not include a linear term in $\theta$. 
Thus internal consistency of the theory implies that the non-linear saturation term of the CoopCM in Eq.~(\ref{eq:ODE_H_2}) follows from the fact that coat invagination starts at a finite $A$, given by Eq.~(\ref{eq:exapnsion_H}). 
We conclude that the CoopCM in fact predicts a flat-to-curved transition at finite coat area, in agreement with recent experimental observations \cite{mund2023}.
Although the geometric motivation is consistent, we note that different microscopic models could lead to the same macroscopic description. 

We still have to clarify at which coat area the invagination process starts. We therefore examine the area of the spherical cap using the expression for the coat radius $R(\theta)=1/H(\theta)$ (Eq.~(\ref{eq:solution_H_2})) in the limit of a flat disk, i.e., $\theta \rightarrow 0$
\begin{equation}
    \lim_{\theta \rightarrow 0} A= 
    \lim_{\theta \rightarrow 0} 2 \pi 
    \left ( \frac{R_0}{\tanh \left( \gamma R_0 
\theta \right)} \right )^2 (1- \cos \theta)=
\frac{\pi}{\gamma^2} \, .
\label{eq:interpretation_gamma}
\end{equation}
We see that the CoopCM indeed automatically takes care of the double limit required to get a flat disc as initial condition.
From Eq.~(\ref{eq:interpretation_gamma}) we also find a second, geometrical interpretation of $\gamma$ as the inverse of the patch radius $R_\mathrm{T}=1/\gamma$ at which the 
invagination of the flat coat starts.
In the limit when the coat approaches maximum invagination, we get 
\begin{align}
    \lim_{\theta \rightarrow \pi } A&= 
    \lim_{\theta \rightarrow \pi } 2 \pi 
    \left ( \frac{R_0}{\tanh \left( \gamma R_0 
\theta \right)} \right )^2 (1- \cos \theta) \nonumber \\
&=
4 \pi R_0^2 \coth^2(\gamma R_0 \pi) \, .
\label{eq:interpretation_R_0}
\end{align}
Thus, the lower bound for the area of the fully invaginated state is $A=4 \pi R_0^2$. 
From the expression, we deduce a second, geometrical interpretation of $R_0$, as the radius of the fully invaginated clathrin coat, which is valid in good approximation if the flat-to-curved transition occurs at finite coat area.
Compared to Eq.~(\ref{eq:ODE_H_2}), Eqs.~(\ref{eq:interpretation_gamma}-\ref{eq:interpretation_R_0}) suggest a complementary perspective on the CoopCM:
The CoopCM interpolates between the finite coat area where the flat-to-curved transition occurs (with circular disk radius $R_T=1/\gamma$) and the fully invaginated spherical coat area (with sphere radius $R_0=1/H_0$) in a purely geometric way, without any additional parameters.

We also note that the CCM is a limiting case of the CoopCM if $\gamma \rightarrow \infty$. 
The CAM is partially recovered from the CoopCM, when we demand  
$ A (\theta=0)= A(\theta=\pi)$, which is true for $\gamma R_0=0.4411$. 
In this case, the corresponding coat area varies by around $10\%$ between $\theta=0$ and $\theta=\pi$ along the domain. 
These considerations suggest that the product $\gamma R_0$ can be taken as a measure which indicates how similar the CoopCM is with respect to the CCM ($\gamma R_0 \gg 1$) and CAM ($\gamma R_0 = 0.4411$).

\begin{figure*}
\centering
\includegraphics[width=.8\textwidth]{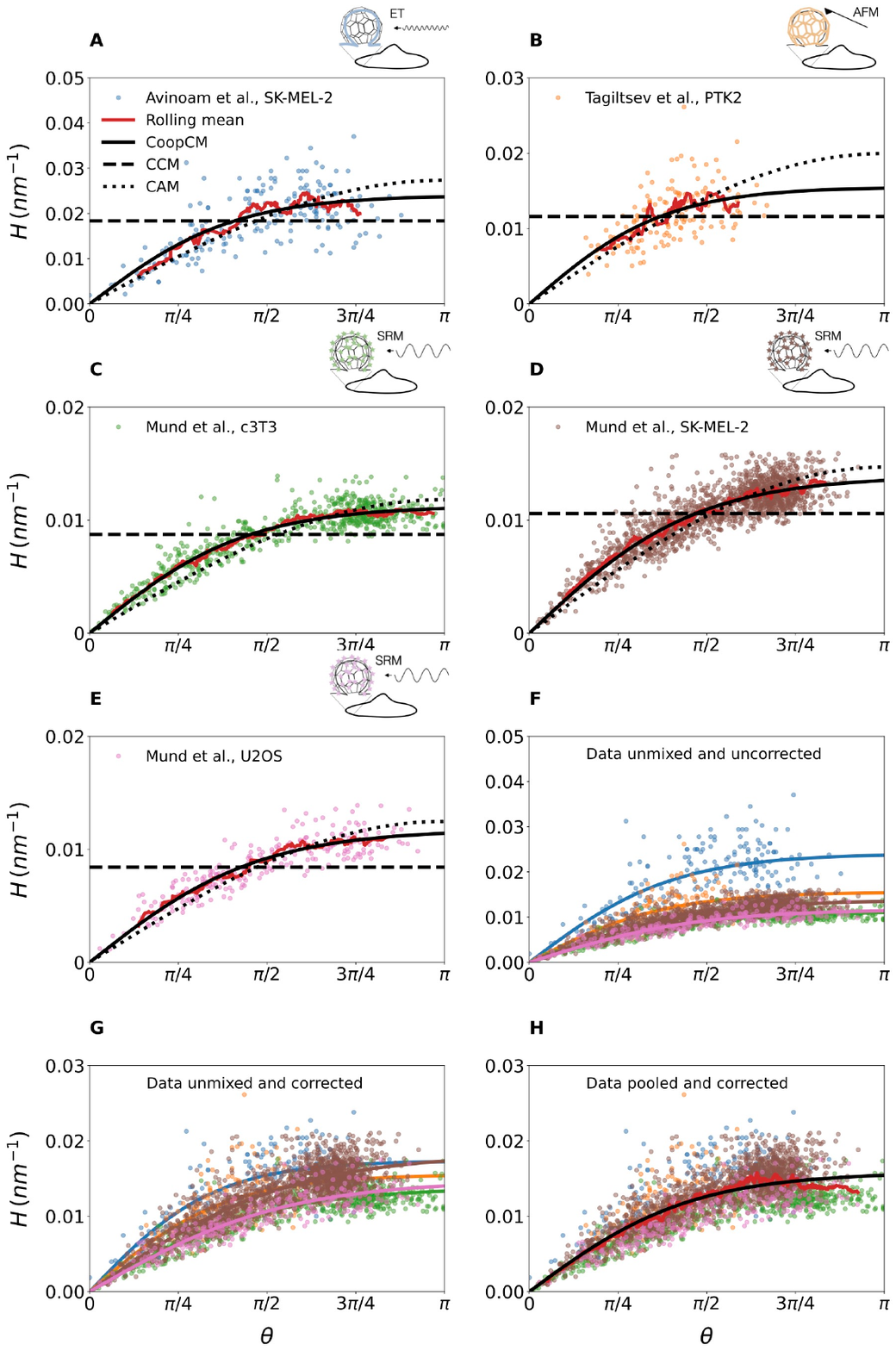}
\caption{\textbf{Experimental data define a consensus pathway that agrees with the CoopCM.} 
(A-H) Curvature $H$ as a function of
invagination angle $\theta$ for different data sets.
The rolling mean (red) and the fit of the CoopCM
(solid black), the CCM (dashed black) and the CAM (dotted black) are shown.
(A) Electron tomography (ET) data \cite{avinoam2015}.
(B) High-speed atomic force microscopy (HS-AFM) data \cite{tagiltsev2021}.
(C-E) Super-resolution microscopy (SRM) data from different cell lines \cite{mund2023}.
(F) Datasets from A-E and the corresponding individual fits of the CoopCM.
(G) Datasets from A-E that are corrected for curvature together with the individual fits of the CoopCM.
(H) Pooled data from the datasets A-E, with the ET and SRM data corrected for curvature.
The results of the fits are documented in Tab.~\ref{table:fitted_parameters}.
}
\label{fig:Fig2}
\end{figure*}

To illustrate the universal pathway of coat invagination as predicted
by the CoopCM, in Fig.~\ref{fig:Fig1} the curvature (B), area (C) and edge length (D) of the clathrin coat are plotted, normalized to their maxima for $H_0=\gamma$. 
Due to the normalization and the fact that we here use $H_0=\gamma$, the shape of the curves for curvature, area and edge length are independent of the specific choice of $H_0$ and $\gamma$.
Apart from the saturation behaviour in curvature, we see that invagination of the clathrin coat starts at a finite initial coat area and therefore also at a finite edge length. 
The edge length has a maximum in the vicinity of the equator.

\begin{table*}
\caption{Parameter values
for fitting the CoopCM (Eq.~(\ref{eq:solution_H_2})) 
to different data sets (cf.~Fig.~\ref{fig:Fig2}).
The fitted values for the datasets of 
Mund et al. \cite{mund2023} agree with 
those reported in \cite{mund2023}.}
\label{table:fitted_parameters}
\begin{tabular}{ccccccc}
\hline
Reference
&
Method
&
Cell line
& $R_0$ ($\unit[]{nm}$)
& $\gamma$ ($\unit[]{nm^{-1}}$) 
& $\gamma R_0$ 
& $A(\theta=0)/A(\theta=\pi) ($\%$)$ \\
\hline
Avinoam \cite{avinoam2015}
&
ET
&
SK-MEL-2
& $\unit[41.6]{}$
& $\unit[0.0189]{}$
& $\unit[0.785]{}$
& $\unit[39.4]{}$
\\
Tagiltsev \cite{tagiltsev2021}
&
HS-AFM
&
PTK2
& $\unit[64.4]{}$
& $\unit[0.0129]{}$
& $\unit[0.834]{}$
& $\unit[35.2]{}$
\\
Mund \cite{mund2023}
&
SRM
&
c3T3
& $\unit[88.8]{}$
& $\unit[0.00817]{}$
& $\unit[0.726]{}$
& $\unit[45.5]{}$
\\
Mund \cite{mund2023}
&
SRM
&
SK-MEL-2
& $\unit[72.0]{}$
& $\unit[0.00942]{}$
& $\unit[0.678]{}$
& $\unit[51.3]{}$
\\
Mund \cite{mund2023}
&
SRM
&
U2OS
& $\unit[85.0]{}$
&$\unit[0.00789]{}$
& $\unit[0.671]{}$
& $\unit[52.4]{}$
\\
Pooled data
&&
& $\unit[63.2]{}$
&$\unit[0.0110]{}$
& $\unit[0.697]{}$
& $\unit[48.9]{}$
\\
\hline
\end{tabular}
\end{table*} 

\section{The consensus pathway of CME}

We now use the CoopCM, in the form of Eq.~(\ref{eq:solution_H_2}), to fit experimental data sets of the coat curvature $H$ as a function of the invagination angle $\theta$ that have been recorded using different experimental methods and different cell lines.
We use ET data \cite{avinoam2015}, HS-AFM data \cite{tagiltsev2021}, and SRM data \cite{mund2023}.
The HS-AFM data is used as kindly provided by the authors \cite{tagiltsev2021}. 

For the ET data, we converted the measured tip radius of membrane pits as a function of $\theta$ to get the coat curvature $H(\theta)$ assuming the geometry of a spherical cap.
Similarly, for the HS-AFM data we converted the measured coat area as a function of the coat radius to obtain $H(\theta)$.
For the SRM data, clathrin-coated pits with negative curvature values and curvature values exceeding a threshold were excluded from the analysis (c3T3: $H>\unit[0.014]{nm^{-1}}$, SK-MEL-2: $H>\unit[0.016]{nm^{-1}}$, U2OS: $H>\unit[0.014]{nm^{-1}}$), similar to what was previously reported (note for U2OS: there was a typo in \cite{mund2023} for the threshold value), because these strongly curved pits lack endocytic marker AP2-GFP and might belong to the Golgi \cite{mund2023}.

The different data sets, the rolling means, and the fits according 
to Eq.~(\ref{eq:solution_H_2}) are shown in 
Fig.~\ref{fig:Fig2}A-E.
For the CoopCM, we determine two parameters from the fit,
$R_0=1/H_0$ and $\gamma$, which are both documented in 
Tab.~\ref{table:fitted_parameters}.
When we compare the results of the fits of the CoopCM to the CCM with $H=C$, where $C$ is a constant, and to the CAM with $H(\theta)=\sqrt{2 \pi(1-\cos \theta)/A_0}$, where $A_0$ is a constant, we find that the CoopCM agrees best with the data for all used imaging techniques and cell lines.
For the CoopCM, $\gamma R_0$ is between 0.67 and 0.83, while the flat-to-curved transition ($A(\theta=0)/A(\theta=\pi)$) occurs at around 35-52$\%$ of the final coat area (Tab.~\ref{table:fitted_parameters}). 
This finding is consistent with the interpretation that the CoopCM predicts coat invagination at finite area, which is in between the CCM ($\gamma R_0 \gg 1$) and the CAM ($\gamma R_0 = 0.4411$).

When we plot the datasets within a single panel, we find that all of them follow similar trends (Fig.~\ref{fig:Fig2}F).
But we also find characteristic differences.
Most importantly, the scale of coat curvature is different for the various datasets.
For SK-MEL-2 cells using ET in Fig.~\ref{fig:Fig2}A,
the curvature values reach around $\unit[0.02]{nm^{-1}}$,
the largest observed values. 
For PTK2 cells using HS-AFM 
in Fig.~\ref{fig:Fig2}B, we observe 
intermediate values of coat curvature of around
$\unit[0.015]{nm^{-1}}$.
For c3T3, SK-MEL-2 and U20S cells using SRM in Fig.~\ref{fig:Fig2}C-E, we find the smallest curvature
values of around $\unit[0.01]{nm^{-1}}$.
Interestingly, the differences between the different cell lines using the same technique in Figs.~\ref{fig:Fig2}C-E are smaller than the differences between different techniques using the same cell line in Figs.~\ref{fig:Fig2}A and D.
The finding can be corroborated when comparing the different values of the preferred coat radius $R_0$ in Tab.~\ref{table:fitted_parameters}.

It is natural to assume that this systematic discrepancy is caused by how the clathrin coats are imaged in the different techniques.
In the ET data of Avinoam et al. \cite{avinoam2015}, the radii of the membranes invaginated by the clathrin coats are measured. 
We therefore expect ET to give large values of curvature.
In the HS-AFM data of Tagiltsev et al. \cite{tagiltsev2021}, the curvature is measured in unroofed cells directly at the clathrin coat.
We thus expect intermediate values of coat curvatures for this techniques.
In the SRM data of Mund et al. \cite{mund2023}, the clathrin coat is labeled by polyclonal antibodies that bind to both clathrin heavy and light chains of permeabilized cells, with a potential preference to the intracellular side due to a higher density of binding sites. 
Therefore, we expect small values of coat curvature. 
These arguments explain why we observe that the values of coat curvature decrease from ET through HS-AFM to SRM (cf.~Tab.~\ref{table:fitted_parameters}).

Since the kinetics of clathrin coat invagination seem to be very similar for all cell lines and techniques (cf.~Figs.~\ref{fig:Fig2}A-F), we decided to pool the data by combining the values from all measurements.
Following our previous reasoning and to exclude any bias in the available data, we corrected the measured data from ET by taking into account the thickness of the clathrin coat and its gap to the membrane. The corrected coat curvature is given by $H=1/(1/H_\mathrm{pit}+h_\mathrm{cc})$, where $H_\mathrm{pit}$ is the measured pit curvature and $h_\mathrm{cc}=\unit[15]{nm}$ is the estimated thickness correction \cite {saleem2015}. 
Similarly, we corrected the measured data from SRM by taking into account the length of the antibodies. 
The corrected coat curvature is then given by  $H=1/(1/H_\mathrm{a}- l_\mathrm{ca}$), where  $H_\mathrm{a}$ is the measured antibody-labeled coat curvature and $l_\mathrm{ca}=\unit[15]{nm}$ is the estimated length of the clathrin antibody.

For the curvature-corrected data, the individual data sets are fitted in Fig.~\ref{fig:Fig2}G. 
The scattering in the experimental data is considerable, which might be partially due to resolution limits.
Moreover, since CME in cells involves many different types of proteins that are all underlying stochastic fluctuations, CME itself is highly stochastic. 
In addition, there are abortive events and there might be subpopulations, both of which are not taken into account here. 
However, we also note that the variability within single data sets is larger than the variability between the fits of the individual data sets, which seems to indicate that CME is rather similar across imaging methods and cell types. 
Thus, we decided to fit and evaluate the pooled data in order to find the mean trajectory for CME (cf.~Fig.~\ref{fig:Fig2}H and Tab.~\ref{table:fitted_parameters}). 
We see good agreement between experimental data and the CoopCM and conclude that the resulting consensus pathway can be used as a reasonable basis for an energetic analysis. 
We also find that the flat-to-curved transition occurs at around $50\%$ of the final coat area.

\section{The energetics of CME}
\label{sec:dynamic_coat}
\subsection{Energy-based model}

To address the question which physical mechanisms dominate 
the invagination of clathrin coated pits, we now
introduce a model that describes the energetics of
the clathrin coat. 
We formulate the total energy of the composite of plasma membrane and clathrin coat by a generalized membrane energy \cite{frey2020b} that includes the most relevant energetic contributions
\begin{align}
\mathcal{H}&= \int_\mathrm{mem} \left[ 2\kappa H^2 +\sigma \right]  d A
\nonumber \\
&+ \int_\mathrm{coat} \left[ -\mu + 2\kappa_c (H-H_\mathrm{c})^2 \right]  d A + \zeta \mathcal{E} \, .
\label{eq:free_energy_general}
\end{align}
The first integral describes the bending and tension energies of the 
plasma membrane, where $\kappa$ is the bending rigidity of the membrane, 
$\sigma$ is the membrane tension and $H$ is the mean
membrane curvature. 
The membrane energy is integrated over both the coated and 
free membrane parts. For the tension energy,
the integral gives the excess area of the membrane, i.e., the additional area compared to the flat state. 

The second integral in Eq.~(\ref{eq:free_energy_general}) describes the polymerization and bending energies of the 
clathrin coat, where $\mu$ is the polymerization energy density, $\kappa_\mathrm{c}$ is 
the coat bending rigidity, and $H_\mathrm{c}$ is the preferred curvature of the coat.
The coat energy is integrated only over the coated membrane parts.
The preferred coat curvature is related to the preferred radius of the clathrin cage
by $H_\mathrm{c}=1/R_\mathrm{c}$.
The last term is the line tension energy, where $\zeta$ is the line tension.
The line tension energy could be due to unsaturated clathrin bonds, additional protein binding at the edge of the clathrin coat or actin pushing on the edges.

In principle, the two integrals should be evaluated over different neutral surfaces, namely
the ones of membrane and coat, respectively, which have a typical distance of \unit[15]{nm} from each other due
to the gap layer of adaptor proteins and the finite thicknesses of the two layers.
In order to formulate a transparent theory in the spirit of thin shell and surface Hamiltonian models, 
here we neglect this effect. This aspect of the model is closely related to the question of 
relevant model parameters; for example, the reported high values for coat rigidity typically apply
for the composite including the gap, and not necessarily for the clathrin coat alone \cite{jin2006}, 
as we will discuss below.

Since it has previously been shown that the energy of the 
free membrane contributes only up to 20\% of the 
whole membrane energy \cite{frey2019}, 
in the following we neglect these contributions
and only consider the membrane of the coated part.
Again we assume that the clathrin coat has 
the shape of a spherical cap.
Moreover, we assume that the properties of the clathrin coat are constant along the whole domain. 
Thus the integration over the coat area in Eq.~(\ref{eq:free_energy_general}) is given by the area $A$. 
The energy of the composite then follows as
\begin{align}
E(\theta)&=4 \pi \kappa \left ( 1 -\cos \theta \right) + \frac{\sigma}{2}
\left(1-\cos \theta  \right) A
\nonumber \\
&-\mu A + \zeta \mathcal{E} +
2 \kappa_c \left(H(\theta)-H_\mathrm{c}\right)^2 A  \, ,
\label{eq:equation2}
\end{align}
where we used the membrane excess area 
$\Delta A=(1-\cos \theta)A/2$.
The first term does not depend on size
due to the conformal invariance of the bending Hamiltonian without spontaneous curvature.
From Fig.~\ref{fig:Fig2}H, we know the consensus pathway of invagination.
Based on this, we can now predict the different energy contributions
in absolute terms by using the coat area $A = 2 \pi R^2 (1-\cos \theta)$ and the coat
edge length $\mathcal{E} = 2 \pi R \sin \theta$ on Eq.~(\ref{eq:equation2}).

\begin{figure*}
\centering
\includegraphics[width=.9\textwidth]{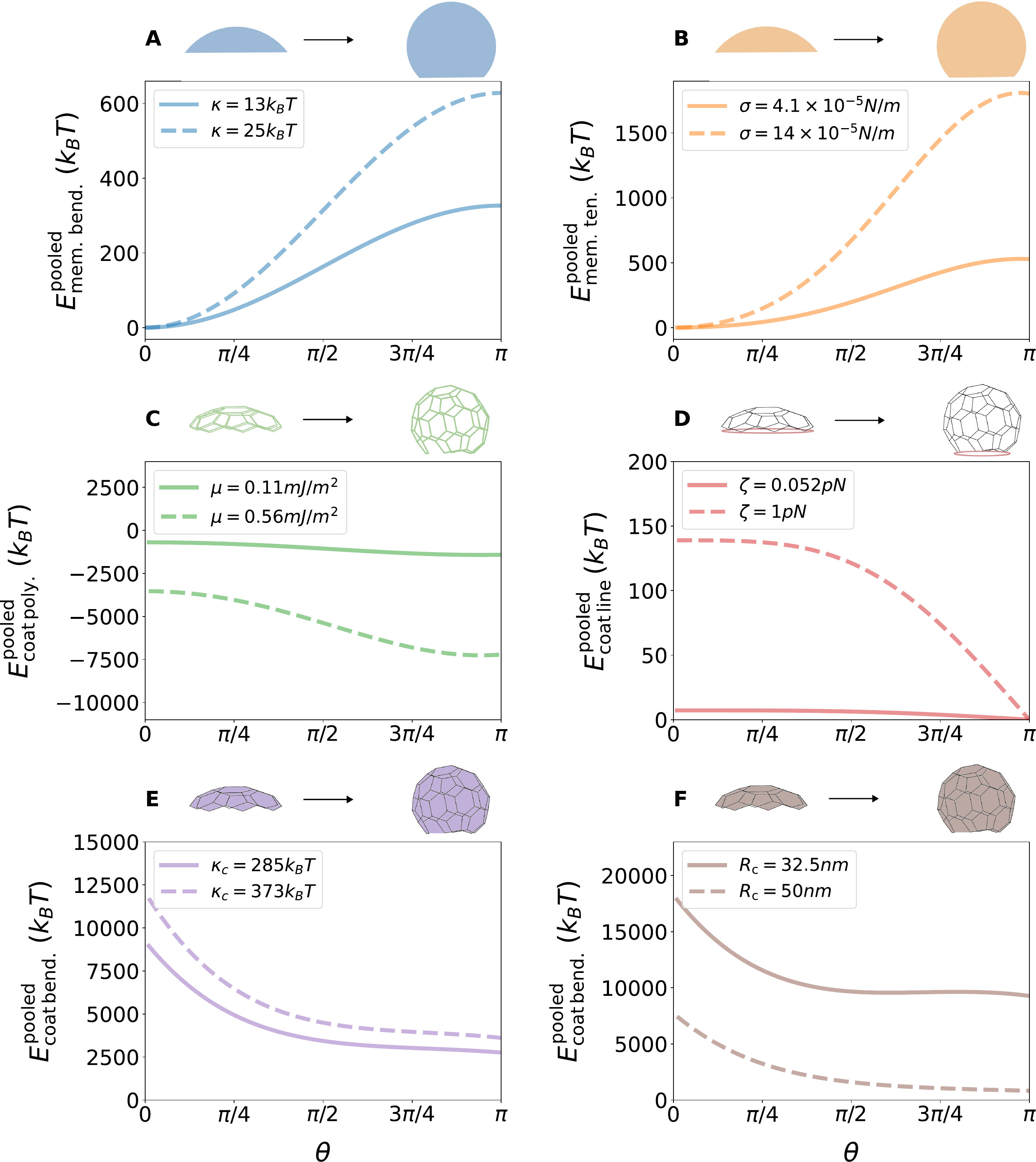}
\caption{\textbf{Different contributions to the energetics
of the consensus pathway.} 
(A) Membrane bending energy.
(B) Membrane tension energy.
(C) Clathrin coat polymerization energy.
(D) Line tension energy of the clathrin coat.
(E and F) Clathrin coat bending energy.
We use the references values given in Tab.~\ref{table:Parameters},
but in addition vary the main quantity of each subplot as indicated
in the legend.
}
\label{fig:Fig3}
\end{figure*}

The results are plotted in Fig.~\ref{fig:Fig3} 
for typical parameters values
as summarized in Tab.~\ref{table:Parameters}. For each subplot,
we use the same reference values, but for the quantity under
consideration, two extreme values are used to demonstrate the
possible variations in energetics.

\begin{table}[b]
\caption{Model parameters. If not
indicated differently, we use the reference value.
NB: The reference
value for the line tension is estimated 
in \cite{lipowsky1992} for a lipid domain in the plasma membrane.
}
\label{table:Parameters}
\begin{tabular}{lll}
\hline
Parameter & Typical range & Reference value \\
\hline
Mem. rigid. $\kappa$ & $\unit[13-25]{k_BT}$ \cite{saleem2015,kumar2016}& 
$\unit[15]{k_BT}$ \cite{tagiltsev2021} \\
Mem. tens. $\sigma$ & $\unit[4.1-14 \cdot 10^{-5}]{N/m}$
\cite{tagiltsev2021,kaplan2020}& 
$\unit[4.1 \cdot 10^{-5}]{N/m}$ \cite{tagiltsev2021}\\
Poly. energ. $\mu$& $\unit[0.11-0.56]{mJ/m^2}$
\cite{saleem2015,tagiltsev2021}& $\unit[0.56]{mJ/m^2}$ \cite{tagiltsev2021}\\ 
Line tens. $\zeta$ & $\unit[0.052-1]{pN}$ 
\cite{saleem2015,lipowsky1992} &  
$\unit[1]{pN}$ \cite{lipowsky1992}\\
Coat rigid. $\kappa_c$ & $\unit[285-373]{k_B T}$ \cite{jin2006,tagiltsev2021} & 
$\unit[373]{k_B T}$ \cite{tagiltsev2021} \\
Cage radius $R_\mathrm{c}$& 
$\unit[32.5-50]{nm}$ \cite{saleem2015,hassinger2017} & 
$\unit[40]{nm}$ \cite{tagiltsev2021} \\
\hline
\end{tabular}
\end{table} 

\begin{figure*}
\centering
\includegraphics[width=.9\textwidth]{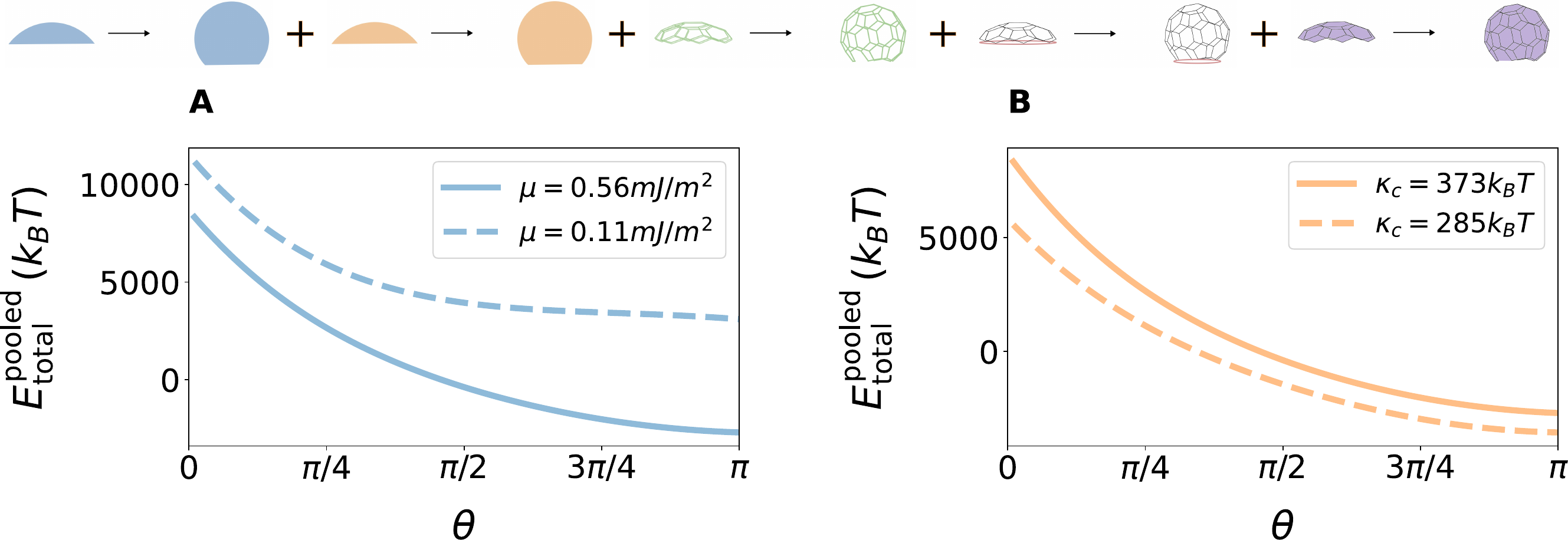}
\caption{\textbf{Total energy for constant model parameters.}
(A) Effect of constant polymerization energy density.
(B) Effect of constant coat rigidity.
The used parameter values are indicated either in the figure legends or in Tab.~\ref{table:Parameters}.
}
\label{fig:Fig4}
\end{figure*}

We first note that both the membrane bending energy (Fig.~\ref{fig:Fig3}A) and the tension
energy (Fig.~\ref{fig:Fig3}B)
monotonically increase as a function of $\theta$.
In contrast, the polymerization energy of the
coat becomes more negative for increasing $\theta$ as
the coat grows in size (Fig.~\ref{fig:Fig3}C),
in agreement with the general notion that release of free energy
by polymerization is the main driving force for invagination. 
Because invagination starts with finite edge length, the line tension 
in Fig.~\ref{fig:Fig3}D is approximately constant
until the coat is halfway invaginated, before it
rapidly decreases over the rest of the time course.
This implies that line tension does not play a driving
role during the initial phases of invagination. Later
it could become important for invagination, as known for
vesicle formation from phase-separated membrane domains 
\cite{Julicher1993,Baumgart2003},
multilayered membranes \cite{baldauf2023}
and sometimes discussed in the context of CME
\cite{lipowsky1992,frey2020b,Noguchi2019},
but our results show that the corresponding energies
are rather small compared to the other
contributions in Fig.~\ref{fig:Fig3}.
By far the largest energetic contribution 
comes from bending the clathrin coat (Fig.~\ref{fig:Fig3}E).
Because $\kappa_c$ has been determined in independent 
experiments, there is no doubt that this value will be very high.
We see that a shallow minimum is caused by the coat approaching its 
preferred curvature, which decreases the energy; 
because at the same time area is still increasing, 
the energy then rises again. 
For large enough values of the preferred cage radius $R_\mathrm{c}$, 
the coat energy monotonically decreases (Fig.~\ref{fig:Fig3}F).
From these plots we conclude that the energetic competition between coat
polymerization energy and coat rigidity strongly dominates the
invagination pathway.

\subsection{Total energy for static coat properties}

Our goal is to predict the dominant physical mechanism for coat 
invagination and we assume that the overall invagination pathway has to be
characterized by a well-defined downhill total energy landscape. 
Moreover, we expect that the flat state $(\theta=0)$ is 
characterized by a zero or negative total energy, so that 
flat growth is possible.
Following our observation that polymerization energy and the stiffness of the coat 
are the dominant factors in coat invagination, we now investigate the effect of these two parameters.
In Figs.~\ref{fig:Fig4}A and B we plot the total energy as a function of $\theta$,
given by Eq.~(\ref{eq:equation2}),
for different polymerization energies $\mu$ 
and coat rigidities $\kappa_\mathrm{c}$, respectively. 
We see that in all considered cases the total energy is positive over a significant part of the domain.
Fig.~\ref{fig:Fig4}A suggests that once started, a large enough polymerization energy could drive the invagination process over the whole lifetime, but that still the initial energy, associated with flat coat assembly, will be positive and hence make flat coat assembly at static parameter values unlikely.
Fig.~\ref{fig:Fig4}B indicates that the coat rigidity always renders the total energy positive for small values of $\theta$.
We conclude that the large positive value of the coat bending energy for small values of $\theta$ makes flat clathrin coat assembly energetically costly and therefore potentially impossible.
Thus the system in the initial stages should rather be characterized by small coat bending
rigidity. On the other hand, in the late stages the coat bending rigidity should
be high in order to enforce the preferred curvature. This suggests that coat bending
rigidity might be time-dependent.

\subsection{Total energy for dynamic coat properties}

Up to now we have seen that constant parameter values for coat polymerization energy, 
line tension, coat rigidity and preferred coat curvature 
do not lead to a negative and monotonically decreasing total energy.
Thus the energetic description of Eq.~(\ref{eq:equation2})
is incomplete and we hypothesize that the clathrin coat is
still plastic during coat invagination. Therefore we now turn to the possibility
that the model parameters have their own dynamics; at the same time, however,
we have to make sure that our CoopCM still remains valid despite these changes.

In order to expand the dynamic description, we 
assume that the invagination pathway of the
coat follows from overdamped dynamics \cite{doi2011}
\begin{equation}
\alpha  \frac{\partial H}{\partial t}  = - \frac{\partial E}{\partial H} \, .
\label{eq:Onsager}
\end{equation}
To make Eq.~(\ref{eq:Onsager}) dimensionally consistent, the friction coefficient
$\alpha$ needs to carry the unit of energy times area and time. 
Since the only relevant energy and area
in our problem are the coat rigidity 
$\kappa_\mathrm{c}$ and the coat area $A$, 
we assume $\alpha \sim \kappa_\mathrm{c} A$.
Moreover, we introduce the invagination rate  $k_\mathrm{i}$ for dimensional reasons with
$\alpha \sim 1/k_\mathrm{i}$.
The assumption makes intuitively sense,
since coat friction increases with 
a stiffer and larger coat that is invaginated at a smaller rate, i.e., over a longer time.
Moreover, we expect that $\alpha \sim \theta$ to incorporate the notion that friction increases during the invagination of the clathrin coat.
In order to simplify Eq.~(\ref{eq:Onsager}), we now make use of the chain rule and obtain
\begin{equation}
\alpha  \frac{\partial H}{\partial \theta}\frac{\partial \theta}{\partial t}    = - \frac{\partial E}{\partial H} \, ,
\label{eq:Onsager_2}
\end{equation}
with $\alpha=\kappa_\mathrm{c} A \theta / k_\mathrm{i}$.
In order to find an expression for $\partial \theta / \partial t$, we use the underlying dynamic assumption of the CoopCM, namely that the area $A$ of the clathrin coat grows along the edge $\mathcal{E}$ by addition of new triskelia with the growth speed  $k_\mathrm{on}$  \cite{mund2023}
\begin{equation}
\frac{\partial A}{\partial t}=k_\mathrm{on} \mathcal{E} \, . 
\label{eq:growth_equation}
\end{equation}
Using once again the assumption that the clathrin coat takes the shape of a spherical cap, we can simplify Eq.~(\ref{eq:growth_equation}) to get
\begin{equation}
2 \dot{R} \tan \frac{\theta}{2} + R \dot{\theta} = k_\mathrm{on} \, .
\label{eq:ODE_theta_R}
\end{equation}
We then use the inverse of Eq.~(\ref{eq:solution_H_2}), 
its derivative and the chain rule to simplify Eq.~(\ref{eq:ODE_theta_R}) which leads to
\begin{equation}
\dot{\theta}=k_\mathrm{on} \frac{1}{2 \frac{\partial R(\theta)}{\partial \theta} 
\tan \frac{\theta}{2}+R(\theta)} \, .
\label{eq:ODE_theta_2}
\end{equation}
After expanding Eq.~(\ref{eq:ODE_theta_2}) up to leading order in 
$\theta$ we find
\begin{equation}
\dot{\theta}=k_\mathrm{g} \frac{1}{\theta}\, ,
\label{eq:ODE_theta_expanded_2}
\end{equation}
with the rate of growth $k_\mathrm{g} =12 \gamma k_\mathrm{on}/(8 \gamma^2 R_0^2-1)$, which carries the unit of $\unit[1]{s^{-1}}$.
Although the expansion in Eq.~(\ref{eq:ODE_theta_expanded_2}) is formally valid only for small values of $\theta$, the functional form is supported by an analysis of the invagination of clathrin coats as a function of pseudotime (the mapped frequency of invagination) \cite{mund2023}. 
The description deviates only for the late stages of coat assembly, where vesicle scission occurs. 
We therefore extrapolate the behavior as stated in Eq.~(\ref{eq:ODE_theta_expanded_2}) to the full domain of $\theta$. 
Now we can put everything together and obtain
\begin{equation}
\frac{\partial H}{\partial \theta}   = - \frac{k_\mathrm{i}}{k_\mathrm{g}} \frac{1}{A \kappa_\mathrm{c}} \frac{\partial E}{\partial H} \, .
\label{eq:Onsager_3}
\end{equation}
In Eq.~(\ref{eq:Onsager_3}) the ratio of the invagination rate and the growth rate $k_\mathrm{i}/k_\mathrm{g}$ defines a number that scales the dynamics. 
Since we have no means to determine those rates from the structural data we analyze, we consider $k_\mathrm{i}/k_\mathrm{g}$ as a free parameter.

\begin{figure*}
\centering
\includegraphics[width=.9\textwidth]{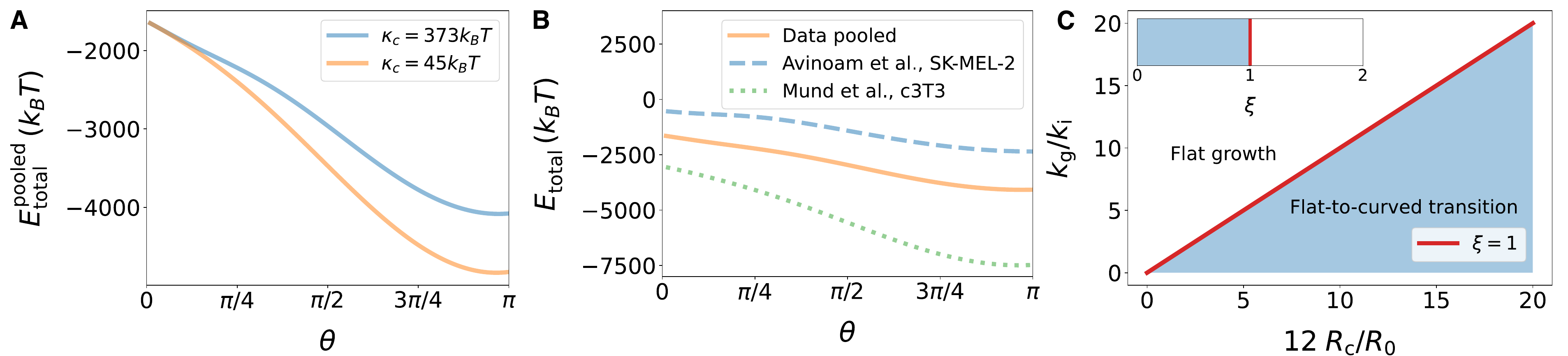}
\caption{\textbf{Total energy for time-varying model parameters}. 
(A) The total energy including all energy contributions as predicted by the energetic model according to Eqs.~(\ref{eq:coat_stiffening_mu}-\ref{eq:coat_stiffening_kappa_c}) using the consensus invagination pathway of Fig.~\ref{fig:Fig2}H.
The used parameter values are indicated either in the figure legend or in Tab.~\ref{table:Parameters}.
(B) The total energy including all energy contributions as predicted by the energetic model
according to Eqs.~(\ref{eq:coat_stiffening_mu}-\ref{eq:coat_stiffening_kappa_c})
using the consensus invagination pathway of Fig.~\ref{fig:Fig2}H,
the pathway with the smallest pits
(cf.~Fig.~\ref{fig:Fig2}A)
and the pathway with the largest pits
(cf.~Fig.~\ref{fig:Fig2}C).
Parameter values used are indicated
in Tab.~\ref{table:Parameters}.
(C) Flat-to-curved transition predicted for
$R_\mathrm{c}$, $R_0$, $k_\mathrm{g}$ and $k_\mathrm{i}$.
Inset: The clathrin coat undergoes a 
flat-to-curved transition if $\xi\le1$.}
\label{fig:Fig5}
\end{figure*}

The particular form of the second term of the prefactor ($1/(A\kappa_\mathrm{c})$) suggests the following assumptions on the right hand side of Eq.~(\ref{eq:Onsager_3}):
When all energy contributions are normalized by $\kappa_\mathrm{c}$, all energy contributions except for the polymerization energy and the coat bending energy are rendered negligible (cf.~Fig.~\ref{fig:Fig3}).
Since the coat polymerization energy and the coat rigidity
are the relevant quantities that drive coat invagination,
we ask under which assumption we can get the
CoopCM from these two terms.
Therefore, we only consider the polymerization and coat energy in
Eq.~(\ref{eq:equation2}) and assume $\mu(H)$ and $\kappa(H)$ 
to depend on the degree of invagination.
From Eq.~(\ref{eq:Onsager_3}) we obtain
\begin{equation}
\frac{\partial H}{\partial \theta}  = 
- \frac{k_\mathrm{i}}{k_\mathrm{g}} \frac{1}{A \kappa_\mathrm{c}}
\frac{\partial}{\partial H} \left(-\mu(H) A 
+2 \kappa_c (H) (H-H_\mathrm{c})^2 A \right ) \, .
\label{eq:ansatz_dynamics}
\end{equation}
Here we assume once again that the
coat area $A$ grows by addition of triskelia 
over the edge of the coat, independent of H.
In order to bring Eq.~(\ref{eq:ansatz_dynamics}) to the same functional form as the CoopCM, given by Eq.~(\ref{eq:ODE_H_2}), and to avoid any internal inconsistency within the model framework, the simplest possible assumption is that the polymerization energy increases linearly with the curvature. Moreover, we must assume that the clathrin coat stiffens with an exponent of $3$. We then have
\begin{align}
\label{eq:coat_stiffening_mu} 
\mu (H) &= \frac{\mu}{2} \left(1+\frac{H}{H_\mathrm{c}} \right)\ ,  \\
\label{eq:coat_stiffening_kappa_c} 
\kappa_\mathrm{c} (H) &= \kappa_\mathrm{c} \left ( 
\frac{H}{H_\mathrm{c}} \right ) ^3 \, . 
\end{align}
The assumption of an initially weak coat seems to be required to allow initial curvature generation. 
Later, the increase in this value reinforces the invagination because it forces the system to adapt to the inherent curvature of the mature lattice.

In principle, the dynamics of CME could also change the preferred
coat curvature $H_\mathrm{c}$. Making $H_\mathrm{c}$ 
dependent on coat curvature would have 
similar effects as making $\kappa_\mathrm{c}$ 
dependent on coat curvature. 
However, all possible sources of coat stiffening such as 
rearrangements in the clathrin coat or
conformational changes in a clathrin triskelion will 
affect $\kappa_\mathrm{c}$, too, 
as it sets the scale of the coat energy.
Therefore, we here consider $\kappa_\mathrm{c}$ to be 
dependent on coat curvature, whereas $H_\mathrm{c}$ is assumed to be fixed
by the geometry of the clathrin triskelion.

The functional form from above is 
the simplest one that can be directly 
linked to the CoopCM.
Using Eqs.~(\ref{eq:coat_stiffening_mu}-\ref{eq:coat_stiffening_kappa_c}) on Eq.~(\ref{eq:ansatz_dynamics}), we get in leading order
\begin{equation}
\frac{\partial H}{\partial \theta}  = \frac{k_\mathrm{i}}{k_\mathrm{g}} 
\frac{\mu}{2\kappa_\mathrm{c} H_\mathrm{c}}
\left ( 1- \frac{12 \kappa_\mathrm{c}}{\mu} H^2 \right)
\, .
\label{eq:pseudo_dynamics}
\end{equation}

By comparing the coefficients in Eq.~(\ref{eq:ODE_H_2}) and Eq.~(\ref{eq:pseudo_dynamics}), we can link the energetic description to the CoopCM and therefore gain a more mechanistic understanding of this initially kinetic model
\begin{align}
H_0&=\sqrt{\frac{\mu}{12 \kappa_\mathrm{c}}}
\, ,
\label{eq:def_H0}
\\
\gamma&= \frac{k_\mathrm{i}}{k_\mathrm{g}}  \frac{\mu}{2 \kappa_\mathrm{c} H_\mathrm{c}}\ . 
\label{eq:def_gamma}
\end{align}
Using the fitted value of $R_0=\unit[63.2]{nm}$ and the polymerization energy $\mu=\unit[0.56]{mJ \, m^{-2}}$, we predict $\kappa_\mathrm{c}=\unit[45]{k_B T}$ from Eq.~(\ref{eq:def_H0}).
Compared to the typical value of coat rigidity in Tab.~\ref{table:Parameters}, our predicted value of $\kappa_\mathrm{c}$ is too small by a factor of 6 to 7, likely due to simplifications in our theory and the assumed functional form of Eqs.~(\ref{eq:coat_stiffening_mu}-\ref{eq:coat_stiffening_kappa_c}).
In particular, a different curvature scale instead of $H_\mathrm{c}$ or higher order terms could contribute to coat stiffening.
However, it is also possible that the experimentally determined value of $\kappa_\mathrm{c}$ is too high, especially because it usually includes the contributions of the mechanics of the gap layer \cite{jin2006}. 
In addition, using the fitted value of $\gamma=\unit[0.0110]{nm^{-1}}$, the coat rigidity of $\kappa_\mathrm{c}=\unit[45]{k_B T}$ and the cage radius of $R_\mathrm{c}=\unit[40]{nm}$ we predict  $k_\mathrm{g}/k_\mathrm{i}=5.5$ from Eq.~(\ref{eq:def_gamma}).
We note that due to the functional form of Eq.~(\ref{eq:Onsager_2}), the dynamics of $\theta$, given by Eq.~(\ref{eq:ODE_theta_2}), will only scale the curvature evolution equation ($\partial H / \partial \theta$). 
Therefore, the connection between the energetic model and the CoopCM remains valid up to a scaling function even if a higher order expansion in $\theta$ in Eq.~(\ref{eq:ODE_theta_expanded_2}) or a different model for coat growth than Eq.~(\ref{eq:growth_equation}) is used.  
In Fig.~\ref{fig:Fig5}A we plot Eq.~(\ref{eq:equation2}) with varying polymerization energy and coat rigidity according to Eqs.~(\ref{eq:coat_stiffening_mu}-\ref{eq:coat_stiffening_kappa_c}) for the parameters in Tab.~\ref{table:Parameters} and $\kappa_\mathrm{c}=\unit[45]{k_B T}$.
For both parameter combinations, the total energy is negative and monotonically decreasing as a function of $\theta$, as it should be in a physical description of CME. 
The result justifies our approach and suggests that polymerization energy and coat rigidity indeed dynamically increase during coat invagination.

We also can check now whether the suggested mechanism is universal in the sense that in all cell lines, the total energy decreases.
We therefore turn again towards the results of the fit of the CoopCM and plot the total energies according to Eqs.~(\ref{eq:equation2}), (\ref{eq:coat_stiffening_mu}) and (\ref{eq:coat_stiffening_kappa_c}) as a function $\theta$ for the parameters that produce the smallest (Fig.~\ref{fig:Fig2}A) and largest pits (Fig.~\ref{fig:Fig2}C). 
The results are shown in (Fig.~\ref{fig:Fig5}B) together with the pooled data for the parameters of Tab.~\ref{table:Parameters}.
Although there are differences between the different cell lines, in all three cases the mechanism of coat stiffening predicts the invagination of the clathrin coat.

We finally predict the invagination pathway of the clathrin coat. 
The flat-to-curved transition occurs when the area of the coat at the transition, $\pi R_\mathrm{T}^2$, is smaller than the final coat area, $4 \pi R_0^2$, i.e., $\pi R_\mathrm{T}^2\le4 \pi R_0^2$. 
Otherwise the clathrin coat grows flat.
From the inequality, we deduce a dimensionless parameter $\xi=R_\mathrm{T}/(2R_0)$ that predicts the flat-to-curved transition if $\xi\le 1$, and flat growth otherwise (Fig.~\ref{fig:Fig5}C, inset).
We connect the invagination radius $R_\mathrm{T}={2 k_\mathrm{g} \kappa_\mathrm{c}}/{(k_\mathrm{i}\mu R_\mathrm{c})}$, to the energetic parameters by using Eq.~(\ref{eq:def_gamma}) and the final coat radius $R_0=\sqrt{{12 \kappa_\mathrm{c}}/{\mu}}$ to the energetic parameters by using Eq.~(\ref{eq:def_H0}).
Using these expressions on the condition of the flat-to-curved transition, we can thus relate the invagination pathway of the clathrin coat to the kinetic parameters,  $k_\mathrm{g}/k_\mathrm{i} \le 12 R_\mathrm{c} / R_0$, which is illustrated in Fig.~\ref{fig:Fig5}C.  
We conclude that the invagination pathway, parameterized by coat growth and invagination is determined by two length scales, the patch radius at which the flat-to-curved transition occurs and the final pit radius.

\section{Discussion}

In this work, we investigated the invagination pathways of clathrin coats from different angles, combining
the analysis of experimental data with modelling.
The analyzed data sets were acquired with
very different methods, namely ET \cite{avinoam2015}, HS-AFM \cite{tagiltsev2021} and SRM \cite{mund2023},
but all of them are high resolution spatial data
that suggest a spherical cap
shape of the clathrin pits and similar
changes in curvature during invagination. 
The spherical cap assumption is further
supported by a detailed analysis of the 
SRM data with a maximum-likelihood model \cite{Wu2023}. 
Since the spherical cap matched the shape of the majority of clathrin coats in the SRM data \cite{mund2023}, and only some sites exhibit asymmetric and irregular deformations, we believe that the spherical cap assumption is well suited to describe the coat shape in our theoretical model.

The geometrical time courses of the invaginations
seem to be fitted well across imaging
techniques and cell types by the recently introduced
cooperative curvature model (CoopCM) \cite{mund2023}, as shown in Fig.~\ref{fig:Fig2}A-F. 
Comparing the trajectories of the different datasets by means of the CoopCM after correcting for experimental uncertainties, we found very similar invagination behavior (Fig.~\ref{fig:Fig2}G). 
We concluded that clathrin coats follow a consensus invagination pathway described by the CoopCM (Fig.~\ref{fig:Fig2}H). 
We note, however, that ultimately the CoopCM is derived phenomenologically and in principle other descriptions might exist, although, to date, we know of no other suitable candidate. 
In particular, a linear growth model was tested before to give worse results \cite{mund2023}. 
The CoopCM suggests that the dynamics of invagination is strongly determined by cooperative generation of curvature in the clathrin lattice. In the future, 
this general conclusion should be validated by
a more microscopic model.

One important result from our model is that a flat-to-curved transition occurs at finite coat area, which agrees with experimental observations \cite{avinoam2015,bucher2018,mund2023}. 
While our model describes the averaged pathway of coat invagination, we cannot rule out that different populations contribute to the analyzed datasets. 
Subpopulations might also follow different pathways and mechanisms for coat invagination. 
Nevertheless, we concluded that the CoopCM defines a consensus pathway which can be used for an energetic analysis.

We then combined the consensus pathway and the CoopCM with an energetic model for the composite of membrane and coat. 
Our main result was that the energy related to coat stiffness dominates all other contributions. 
Moreover, within the framework of our interface model and assuming no further energetic contributions other than those of the membrane and the clathrin coat, our analysis suggests that the coat stiffness changes during coat invagination.
When using a constant value for stiffness and taking the high values measured for pits in their late stages, our model predicts that coat invagination is energetically unfavorable. 
Our model therefore predicts that coat stiffness starts at low values and dynamically increases during invagination, in agreement with the emerging notion that clathrin coats are plastic and cooperative.
Although there is no experimental evidence to date, we point out that additional unreported molecular mechanisms may drive clathrin coat invagination.

We showed that the CoopCM can be derived from the energetic model when assuming that the coat polymerization energy increases linearly with coat curvature and the clathrin coat stiffens with a power law exponent of three with coat curvature. 
We note that a power law exponent of three also relates bending stiffness and thickness of thin sheets \cite{deserno2015}, suggesting that effective thickness might grow linear with curvature. 
In practise, this viewpoint of continuum mechanics might however be too naive, because clathrin lattices have very specific discrete architectures. 
Rather, this prediction of our theory should be tested experimentally.
Our prediction that the polymerization energy of the clathrin coat increases linearly with the curvature of the coat is consistent with experiments and theory predicting that clathrin assembly is stabilized on curved surfaces and that clathrin triskelia in curved coats contain more energy than in flat clathrin lattices \cite{zeno2021,guo2022}.
In fact, the possibility of a curvature-dependent polymerization energy has been discussed previously \cite{kumar2016}.
Possible driving factors for coat stiffening and increase of the effective thickness on a more microscopic level are accumulation of phosphorylated clathrin light chains \cite{maib_cargo_2018}, conformational changes in clathrin triskelia \cite{obashi_conformational_2023}, rearrangements within the coat or mending of lattice defects \cite{sochacki2020}, filling up lattice vacancies \cite{frey2020}, release of elastic energy within the coat \cite{tagiltsev2021}, increasing the density of the clathrin coat \cite{sochacki2017}, solidification of the coat \cite{lipowsky1992,cordella2015}, changing clathrin to AP2 adapter ratio during coat assembly \cite{bucher2018}, the organized binding of clathrin triskelia to adaptor clusters \cite{Paraan2020}, mechanics of the coupling between membrane and clathrin coat through adapter proteins \cite{jin2006}, and the stabilization of membrane curvature by clathrin in a ratchet-like manner \cite{Cail2022}.

All these mechanisms require flexibility and weak interactions within the coat during invagination initially.
Lattice rearrangements seem plausible,
given that the coat shows triskelia exchange 
in the flat state \cite{wu2001,avinoam2015} 
but also can disassemble fast once it is complete \cite{smith2022}.
The notion of weak interactions 
is supported by the fact that
the legs of clathrin triskelia are flexible and 
bind only weakly to each other \cite{morris2019}.
Moreover, clathrin triskelia are not perfectly aligned with 
respect to each other even in curved configurations \cite{fotin2004}.

The stiffening of the clathrin coat could be complemented by an increasing line tension $\zeta$, for example by polymerizing actin pushing from the periphery of the coat \cite{{serwas2022,yang2022}}.
However, our data also shows that this might be dispensable under normal conditions, in line with previous experiments \cite{boulant2011}.
Moreover, the scenario would still require that the clathrin coat stiffens during invagination, because otherwise flat clathrin coat assembly cannot be explained. 
However, analyzing this scenario is beyond the scope of our model and future work will therefore be required. 
Our model is also not suited to analyze the invagination of clathrin coats that do not assume a spherical shape as for example in yeast \cite{Mund2018}.

Our interface model describes the mean invagination pathway of the clathrin coat and does not consider deviations from the spherical cap shape or clathrin lattices that are not regularly assembled. 
It will require future work and more microscopic models to dissect whether these aspects can impact the energetics of coat invagination and how coat stiffening occurs microscopically.
In principle coat stiffening could happen in a continuous way,
similar to releasing elastic energy within the coat
\cite{tagiltsev2021}, or in a discrete or ratchet-like
way, for example due to conformational changes
in the clathrin triskelia, rearrangements
within the clathrin coat \cite{obashi_conformational_2023,sochacki2020},
or by filling up lattice vacanices \cite{frey2020}. 
Therefore, experimentally the CoopCM could be tested by studying the thickness, the density or the mobility of single triskelia within the clathrin coat as a function of time.

The notion of coat stiffening might be related
to the ongoing discussion whether clathrin 
can generate membrane curvature alone \cite{kuey2022}
or whether it is only stabilising membrane curvature generated by adaptors and accessory protein 
\cite{Cail2022,stachowiak2022}.
If membrane curvature was generated mostly by adaptor proteins, clathrin could stabilize the constant
membrane curvature and the clathrin coat would
immediately grow in a stiff configuration.
However, if membrane curvature was mostly
generated by clathrin, the coat curvature would
increase during assembly because the clathrin coat
stiffens. 
We speculate that coat stiffening occurs when
the ability to generate membrane curvature 
by other proteins is insufficient.
If so, clathrin assembly and
coat stiffening could then drive membrane 
invagination.
If coat stiffening is dispensable because
coat curvature is generated by other proteins,
this notion could also explain 
the seemingly conflicting experimental findings
that report that clathrin coat invagination
follows a model different from the
flat-to-curved model
\cite{Scott2018,willy2021,nawara2022}.

To conclude, our results imply 
that the clathrin coat is more plastic during coat
invagination than formerly appreciated and dynamically stiffens
during coat invagination.

\section*{Acknowledgements}

We thank Markus Mund, Aline Tschanz and Jonas Ries 
for helpful discussions and a critical reading of the manuscript.
We also kindly acknowledge Simon Scheuring for providing the HS-AFM data for the analysis of clathrin coat invagination.
We thank the reviewers of previous versions
of this manuscript for useful feedback that helped
us to improve this work.
F.F. acknowledges financial support by the
NOMIS foundation. U.S.S. was supported by the Deutsche Forschungsgemeinschaft (DFG, German Research Foundation) under project number 240245660 (SFB 1129).
Moreover he is a member of the Interdisciplinary Center for Scientific Computing (IWR) at Heidelberg.


%

\end{document}